\UseRawInputEncoding
\documentclass[titlepage,preprint,amsmath,amssymb,aps,superscriptaddress]{revtex4-2}

\usepackage{graphicx}
\usepackage{hyperref}
\usepackage{bm}
\usepackage{dcolumn}
\usepackage{tabstackengine}
\usepackage[dvipsnames]{xcolor}
\usepackage{ulem}
\stackMath{}
\usepackage{subcaption}

\begin{document}

\title{Modelling the evolution of an ice sheet's weathering crust}

\author{Tilly Woods}\email{woods@maths.ox.ac.uk}\affiliation{Mathematical Institute, University of Oxford, Oxford, UK}
\author{Ian J. Hewitt}\affiliation{Mathematical Institute, University of Oxford, Oxford, UK}

\begin{abstract}
The weathering crust is a layer of porous ice that can form at the surface of an ice sheet. It grows and decays in response changing weather and climate conditions, affecting the albedo, the melt rate, and the transport of meltwater across the surface. To understand this behaviour, we seek time-dependent solutions to a continuum, thermodynamic model for the porosity, temperature and thickness of the weathering crust, and the internal and surface melt rates. We find solutions using a numerical enthalpy method, presented in this study. We use idealised `switching' and sinusoidal forcings to explore the different dynamics exhibited during growth and decay, the timescales involved, and the impact of diurnal vs. annual variations. The results demonstrate qualitative agreement with observations, and provide insight into the relative importance of different surface heat fluxes during the growth and decay of the crust. The model therefore provides a useful tool for exploring the response of the weathering crust to climate change.
\end{abstract}


\date{\today}

\maketitle

\section{Introduction}
The weathering crust is a layer of porous ice that forms at the surface of ice sheets and glaciers \citep{Muller1969,Cooper2018}. The structure is formed by shortwave radiation from the sun penetrating below the surface of the ice, where it acts as an internal heat source, causing internal melting \citep[][]{Brandt1993,Cooper2021}. Weathering crusts occur on bare (snow-free) ice in the ablation zone \citep{Cuffey2010}, where there is a net loss of ice mass at the surface, for example on the south-western Greenland ice sheet \citep[e.g.][]{vandenBroeke2008}. The weathering crust is on the order of 1 m thick, with impermeable ice below, and is mostly saturated with meltwater \citep{Cooper2018,Cook2016,Irvine-Fynn2021}.

The weathering crust is of interest for multiple reasons. Firstly, it affects the transport of meltwater across the surface of the ice sheet \citep[][]{Cooper2018,Irvine-Fynn2021}, into nearby surface streams.
This in turn impacts the ice sheet mass loss and consequent effect on sea level \citep[][]{Andersen2015}. For example, surface meltwater can be transiently stored or refrozen in the weathering crust \citep{Smith2017,Cooper2018}, and it can be routed to the bed of the ice sheet via moulins \citep[][]{Smith2017}, where it affects the ice sheet motion \citep[][]{Bell2008,vandeWal2008} before being expelled downstream (e.g. into the ocean) through a subglacial drainage network \citep[e.g.][]{Chandler2013}. Moreover, microbial biomass is transported in the surface meltwater \citep{Cook2016,Irvine-Fynn2021}, affecting the provision of nutrients to downstream environments \citep{Stibal2012} and the carbon fluxes between the ice sheet and the atmosphere \citep{Cook2012}.

Secondly, the appearance and structure of the weathering crust affect how much radiation is reflected or absorbed by the ice \citep{Tedstone2020}, which in turn affects how much the ice melts. For example, a weathering crust reflects more radiation than the glazed, bare ice that is left behind when the weathering crust is removed \citep{Tedstone2020}. The reflectivity of a surface is quantified by the albedo -- the proportion of outgoing to incoming shortwave radiation. Darker surfaces have a lower albedo and absorb more radiation, leading to increased melt.  In addition to the ice structure, microbes (e.g. algae) and impurities within the weathering crust interact with the ice structure \citep{Tedstone2020,McCutcheon2021} and act to lower the albedo \citep[][]{Benning2014,Tedesco2016,Hotaling2021}.

The weathering crust is a naturally transient structure. It only forms in the summer months when there is enough shortwave radiation to cause internal (subsurface) melting \citep[e.g.][]{Cook2016}. On a shorter timescale, the weathering crust responds to rapidly changing weather conditions, growing and decaying on the order of hours and days \citep[e.g.][]{Schuster2001}. Clear, sunny conditions favour formation of the crust, whereas warm, windy, cloudy conditions favour removal \citep{Muller1969,Schuster2001}.

In order to understand the behaviour of the weathering crust and what impact it has on melting and the transport of meltwater, we need to understand its transient response to changing weather and climate conditions. Existing studies have primarily focussed on understanding the weathering crust through observations from the field \citep[e.g.][]{Muller1969,Cooper2018, Cook2016}. Until now, there have been limited attempts to model the weathering crust \citep{Schuster2001,Hoffman2008, Hoffman2014,Woods2023}, and these have tended to focus on specific aspects. \citet{Schuster2001,Hoffman2008, Hoffman2014,Woods2023} all present one-dimensional (vertical) models forced by the surface energy balance components (e.g. shortwave and longwave radiation and turbulent heat fluxes). \citet{Schuster2001} presented a physically-based 6-layer model for the density of the weathering crust, with density changes driven by internal melting resulting from internal shortwave radiation. \citet{Hoffman2008,Hoffman2014} modelled the temperature within the ice using the heat equation with a heat source provided by internal shortwave radiation, gaining predictions of surface lowering and internal melt extent over time. The more complex continuum model we presented in \citet{Woods2023} models the surface lowering, internal and surface melt, weathering crust porosity and thickness, and internal ice temperature simultaneously, based on mass conservation, energy conservation, and internal shortwave radiation. That study focussed only on steady solutions. However, studying the full time-dependent problem is crucial for extending our understanding of the physical processes governing weathering crust evolution, and exploring the response of the weathering crust, melting and runoff to climate change.

In the present study, we extend our work from \citet{Woods2023} to seek time-dependent solutions of the weathering crust model presented in that study. We begin in Section \ref{sec:model} by summarising the model and presenting the model equations relevant for this work. We then present a numerical enthalpy method for solving the time-dependent equations and summarise relevant background in Sections \ref{sec:numerical_solution} and \ref{sec:background_info}. In Section \ref{sec:switching} we consider idealised `switching' experiments to gain insight into the physical processes controlling the growth and decay of the weathering crust. In Section \ref{sec:periodic} we use sinusoidal forcings to explore behaviour closer to the reality. Finally, we discuss our findings, avenues for future work, and conclude in Section \ref{sec:conclusion}.

\section{Time-dependent model}
\label{sec:model}

\begin{figure}
    \centering
    \includegraphics[width=0.9\textwidth]{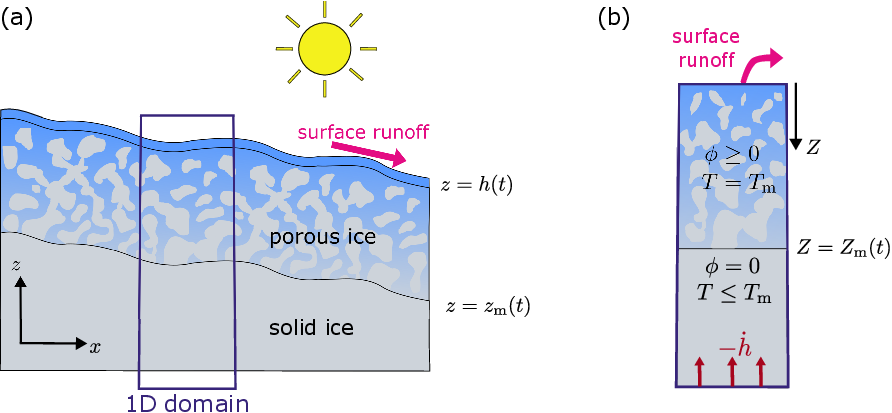}
    \caption{\scriptsize The setup for the weathering crust model in (a) the general two-dimensional case and (b) the one-dimensional case. In the simplest case, the domain consists of two regions: a porous region (the weathering crust) above a cold, solid ice region. Variables are defined within the main text. In the one-dimensional domain, we transform coordinates to $Z = h(t)-z$, meaning that the ice surface is fixed and the cold ice below gets advected towards to surface at rate $-\dot{h}$. Any surface meltwater is removed instantaneously.}
    \label{fig:setup_schematic}
\end{figure}

In \citet{Woods2023}, we presented a fully three-dimensional continuum model for the evolution of the weathering crust, accounting for mass conservation, energy conservation and internal shortwave radiation. We model the porous weathering crust and the solid ice below it, as shown in Figure \ref{fig:setup_schematic}. Here, we present a simplified version of that model under the following assumptions. We restrict ourselves to one (vertical) dimension, and assume that the weathering crust is completely saturated with meltwater. Given that there is always a layer of impermeable ice beneath the weathering crust, this results in no flow of meltwater.

The volume fraction of the water-filled pores is given by the porosity $\phi$, and the volume fraction of ice is $1-\phi$. The temperature $T$ is assumed to be homogeneous at the pore scale, and we assume local thermal equilibrium, meaning that if both ice and the water are present (i.e. in the porous weathering crust) then both phases are at the melting temperature $T_\mathrm{m}$. In regions of pure ice ($\phi=0$), we must have $T \le T_\mathrm{m}$, and in regions of pure water ($\phi=1$), $T \ge T_\mathrm{m}$. Furthermore, we assume that the ice and water have shared properties, such as density $\rho$, thermal conductivity $k$ and specific heat capacity $c$.

The ice surface (at the interface with the air) is at $z=h(t)$, which evolves due to surface melting (see kinematic condition and surface energy balance below). This melting and subsequent surface lowering produce a thin layer of pure meltwater above the surface, which we do not model explicitly. It is assumed that all such surface meltwater immediately runs off laterally into nearby surface streams. As done in \cite{Woods2023}, we transform coordinates to $Z = h(t)-z$, the depth below the ice surface. The equations are presented using the $Z$ coordinate, where the ice surface is now at $Z=0$. The model setup is shown in Figure \ref{fig:setup_schematic}.

\subsection{Mass conservation}
The mass conservation equations for the water and ice reduce to
\begin{equation}
\label{eq:mass_cons}
\rho \Bigg( \frac{\partial \phi}{\partial t} + \dot{h} \frac{\partial \phi}{\partial Z} \Bigg) = m_\mathrm{int},
\end{equation}
where $m_\mathrm{int}$ is the internal melt rate, and $\dot{h}$ ($\leq 0$) is the rate of change of surface height. The second term on the left-hand side accounts for advection of ice towards the surface. In the physical $z$-frame, the ice is stationary and melting causes the surface $z=h(t)$ to lower at rate $-\dot{h}$. In the $Z$-frame, in which the ice surface is fixed at $Z=0$, this manifests itself as the ice advecting backwards towards $Z=0$.

\subsection{Energy equation}
The energy equation is
\begin{equation}
\label{eq:temp_eqn}
\rho c \Bigg( \frac{\partial T}{\partial t} + \dot{h} \frac{\partial T}{\partial Z} \Bigg) = k \frac{\partial^2 T}{\partial Z^2} - \frac{\partial F}{\partial Z} - \mathcal{L} m_\mathrm{int} ,
\end{equation}
where $F$ is the net downwelling internal shortwave radiation (discussed below), and $\mathcal{L}$ is the latent heat of fusion for water. The second term on the left-hand side again accounts for advection of ice towards the top of the domain $Z=0$. The terms of the right-hand side represent the conduction of heat following Fourier's law, the internal heat source provided by the internal shortwave radiation $F$, and the energy sink associated with melting the ice. (This can also be a source, when $m_\mathrm{int}<0$, which corresponds to refreezing).

\subsection{Internal shortwave radiation}
Shortwave radiation from the sun - in particular its penetration below the ice surface - is a key ingredient for forming a weathering crust. Some shortwave radiation penetrates below the surface and is absorbed internally, acting as an internal heat source and producing the porous structure of the weathering crust. Some is absorbed at the surface, and some is reflected back into the atmosphere. In our model, we partition the absorbed shortwave radiation between the surface and the subsurface using a parameter $\chi$ \citep{Hoffman2014,Law2020,Woods2023}.

The total amount of shortwave radiation absorbed is $(1-a)Q_\mathrm{si}$, where $Q_\mathrm{si}$ is the incoming shortwave radiation and $a$ is the albedo (the ratio of outgoing to incoming shortwave radiation). Of that, $\chi(1-a)Q_\mathrm{si}$ is absorbed at the surface and $(1-\chi)(1-a)Q_\mathrm{si}$ is absorbed internally, distributed over depth, as determined by the flux term $F$, which we now describe.

Both downwards and upwards internal radiation can be modelled explicitly using a two-stream approximation, which accounts for the absorption and scattering within the ice \citep[e.g.][]{Perovich1990,Liston1999,Taylor2004,Taylor2005,Woods2023}. If the absorption and scattering coefficients are assumed uniform, as we do here, they can be solved for analytically, and the result is that the \emph{net} downwards radiation $F$ is given by
\begin{equation}
\label{eq:F}
    F = (1-\chi) (1-a) Q_\mathrm{si} e^{-\kappa Z},
\end{equation}
where $\kappa = \sqrt{\alpha^2 + 2 \alpha r}$ is the extinction coefficient, related to the absorption coefficient $\alpha$ and the scattering coefficient $r$. In this study, we also assume $\alpha$ and $r$, and consequently $\kappa$, to be constant in time. The internal shortwave radiation, and the internal heat source it provides, decay exponentially with depth (this is sometimes referred to as Beer's law). The explicit modelling of the upwelling radiation coming back out of the ice surface also determines a relationship between the absorption and scattering coefficients and the albedo \citep{Woods2023}; that is
\begin{equation}
    a = \frac{(1-\chi)(\kappa - \alpha)}{(1-\chi) \kappa + (1+\chi) \alpha}.
\end{equation}

\subsection{Kinematic condition}
The kinematic condition for the ice at the ice surface $Z=0$ is
\begin{equation}
\label{eq:kinematic_ice}
-\rho (1-\phi)  \dot{h} = \rho m_\mathrm{surf} \quad \mathrm{at} \quad Z=0,
\end{equation}
where $m_\mathrm{surf}$ is the surface melt rate. We assume that any excess water at the ice surface runs off instantaneously (rather than allowing it to pool on the surface); combining the rate of surface melting $m_\mathrm{surf}$ with the rate $-\phi \dot{h}$ at which already-liquid water is released from the lowering surface, this means that the runoff occurs at rate $r_\mathrm{off} = - \dot{h}$.

\subsection{Surface energy balance}
The surface energy balance is written (see \citet{Woods2023}) as
\begin{equation}
\label{eq:SEB_simplified}
k \frac{\partial T}{\partial Z} + \chi (1-a) Q_{\mathrm{si}} + Q_0- \upsilon (T - T_\mathrm{m}) = \rho \mathcal{L} m_\mathrm{surf} \quad \mathrm{at} \quad Z=0,
\end{equation}
where $Q_\mathrm{si}$ is the incoming shortwave radiation, and $Q_0 - \upsilon(T-T_\mathrm{m})$ is a linear approximation to the net incoming longwave radiation and turbulent heat flux. It is possible to calculate $Q_\mathrm{si}$, $Q_0$ and $\upsilon$ from observed weather data (see \citet{Woods2023}). In this study, however, we prescribe $Q_\mathrm{si}$ and $Q_0$ as idealised, time-dependent forcings, and take $\upsilon$ to be constant. 

The way that the surface energy balance is interpreted depends on whether or not the surface is melting. When the surface is melting, the surface energy balance determines the surface melt rate $m_\mathrm{surf}>0$, and the boundary condition for the temperature is $T=T_\mathrm{m}$ at $Z=0$. When the surface is not melting, $m_\mathrm{surf}=0$ and the surface energy balance provides the boundary condition for the temperature $T \leq T_\mathrm{m}$ at $Z=0$.

\subsection{Far field boundary conditions}
We prescribe that the ice is cold at depth,
\begin{equation}
\label{eq:Tinf_BC}
    T \to T_\infty \quad \mathrm{as} \quad Z \to \infty,
\end{equation}
where $T_\infty<T_\mathrm{m}$ is the deep temperature of the ice. Consistent with previous assumptions, this also means that it is solid ice, $\phi=0$.

\section{Enthalpy method}
\label{sec:numerical_solution}
We are interested in solutions of the time-dependent model presented in Section \ref{sec:model}. To find these, we reframe the problem in terms of enthalpy, and solve it numerically.

There are three different types of regions we could have in the model domain: (1) cold regions of only ice ($T<T_\mathrm{m}$, $\phi=0$), (2) temperate regions with a mixture of ice and water ($T=T_\mathrm{m}$, $0<\phi<1$), and (3) warm regions of only water ($T \ge T_\mathrm{m}$, $\phi=1$).
We know the porosity in the only-water and only-ice regions, and we know the temperature in the ice-and-water regions. Hence, in each of the three regions, we only need to solve for one of $\phi$ and $T$: we solve the mass conservation equation \eqref{eq:mass_cons} for $\phi$ in ice-and-water regions, and solve the energy equation \eqref{eq:temp_eqn} for $T$ in the only-ice and only-water regions. To implement this directly would require tracking the free-boundaries between neighbouring regions and solving different equations in different parts of the domain.

The enthalpy method \citep[][]{Alexiades1993,Meyer2017,Voller1981AccurateMethod,Ashwanden2012,Hewitt2017} is an alternative approach which we choose to use here. It removes the need to track the free boundaries between the three different types of regions, and allows us to solve a single equation for the enthalpy $H$ in the whole domain. This also has the advantage of accommodating situations in which there are multiple free boundaries, and in which these boundaries come and go, automatically.

\begin{figure}
    \centering
    \includegraphics[width=0.8\textwidth]{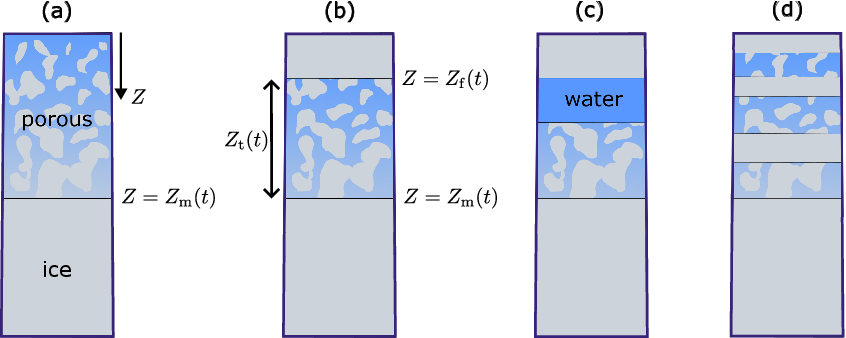}
    \caption{\scriptsize Possible transient configurations of the weathering crust model. Configurations (a) and (b) are encountered in this study. Configurations (c) and (d) are  other, more general, possibilities.}
    \label{fig:transient_config_schematic}
\end{figure}

The enthalpy is defined as
\begin{equation}
    H = \rho c (T-T_\mathrm{m}) + \rho \mathcal{L} \phi,
\end{equation}
i.e. the sum of the sensible and latent heat. The porosity $\phi$ and temperature $T$ can be recovered from the enthalpy $H$ as follows:
\begin{equation}
\label{eq:T_phi_from_H}
\begin{split}
&\phi = \begin{cases}
    0, & H \leq 0,\\
    \frac{H}{\rho \mathcal{L}}, & 0<H<\rho \mathcal{L},\\
    1, & H \ge \rho \mathcal{L},
\end{cases} \\
& T = \begin{cases}
    T_\mathrm{m} + \frac{H}{\rho c}, & H \leq 0, \\
    T_\mathrm{m}, & 0<H<\rho \mathcal{L},\\
    T_\mathrm{m}+ \frac{H}{\rho c} - \frac{\mathcal{L}}{c}, & H \ge \rho \mathcal{L}.
\end{cases}
\end{split}
\end{equation}
The three different cases correspond to there being only ice ($H \leq 0$), both ice and water ($0<H<\rho \mathcal{L}$), and only water ($H \geq \rho \mathcal{L}$). The first holds in the cold, solid ice beneath the weathering crust, and the second holds in the weathering crust. In the simplest weathering crust setup (as shown in Figure \ref{fig:setup_schematic}), we have only these two regions. In general, we could have a combination of only-ice, only-water and ice-and-water layers (such as the examples shown in Figure \ref{fig:transient_config_schematic}). The enthalpy method means that a more complicated situation like this should be no more difficult to solve than the simple two layer case.

By combining the mass conservation equation \eqref{eq:mass_cons} and the energy equation \eqref{eq:temp_eqn} to eliminate the internal melt rate $m_\mathrm{int}$, we get a single equation for the enthalpy,
\begin{equation}
\label{eq:enthalpy_eqn}
\frac{\partial H}{\partial t} + \dot{h} \frac{\partial H}{\partial Z} = k \frac{\partial^2 T}{\partial Z^2} - \frac{\partial F}{\partial Z},
\end{equation}
which is solved everywhere in the domain. Once we have solved the enthalpy equation \eqref{eq:enthalpy_eqn} for $H$, we can recover $T$ and $\phi$ using \eqref{eq:T_phi_from_H}. The boundaries between ice-only and ice-and-water region are identified as where $H=0$, and boundaries of water-only regions are identified by $H=\rho \mathcal{L}$.

Details of the numerical implementation of the enthalpy method are found in Appendix \ref{sec:enthalpy_app}. This includes a description of how we decide which of the two different cases (melt or no-melt) of the surface energy balance \eqref{eq:SEB_simplified} to use in an implicit timestepping setting, and how to deal with the diffusion term in the enthalpy equation \eqref{eq:enthalpy_eqn}.

\section{Solution preliminaries}
\label{sec:background_info}
In Sections \ref{sec:switching} and \ref{sec:periodic} below, we use solutions to the time-dependent model presented in Section \ref{sec:model} to understand the physical processes involved in the growth and decay of the weathering crust. Before doing this, we discuss how the forcings in the model relate to the real world, and we formulate the steady-state solutions.

\subsection{Connecting model and observations}
It is qualitatively observed that clear, sunny conditions favour formation of the weathering crust, whereas warm, windy, cloudy conditions favour removal \citep{Muller1969,Schuster2001,Tedstone2020}.
These conditions can be related to the forcings $Q_\mathrm{si}$ and $Q_0$ used in our model. $Q_\mathrm{si}$ is the shortwave radiation (direct radiation from the sun), which is high in sunny conditions and low in cloudy or dark conditions. It is always positive. $Q_0$ is a combination of longwave radiation and turbulent heat fluxes \citep{Woods2023}, and can be positive or negative. Turbulent heat fluxes account for the fact that, if the air is warmer than the surface, winds can cause turbulent eddies which transfer heat from the air to the surface \citep{Cuffey2010}. The turbulent heat flux is greater when the wind is stronger and the air temperature is warmer. Therefore, $Q_0$ is larger in warm, windy conditions. Increased cloud cover also leads to an increase in $Q_0$, since longwave radiation emitted by the Earth gets reflected back down to the surface in the presence of clouds. Hence, based on observations \citep{Muller1969,Schuster2001}, we expect that large $Q_\mathrm{si}$ favours weathering crust formation, and large, positive $Q_0$ favours weathering crust removal.

Note that negative $Q_0$ corresponds to conditions, such as cold air temperatures, in which the surface would freeze in the absence of shortwave radiation. The heat flux $Q_0$ is completely absorbed at the surface, whereas the absorption of $Q_\mathrm{si}$ is split between the surface and the subsurface (with the proportion given by the parameter $\chi$). Therefore, a positive value of $Q_0$ will result in surface melting, and sufficiently negative value of $Q_0$ (compared to $Q_\mathrm{si}$) will result in surface freezing.

\subsection{Steadily melting solutions}
\begin{figure}
    \centering
    \includegraphics[width=\textwidth]{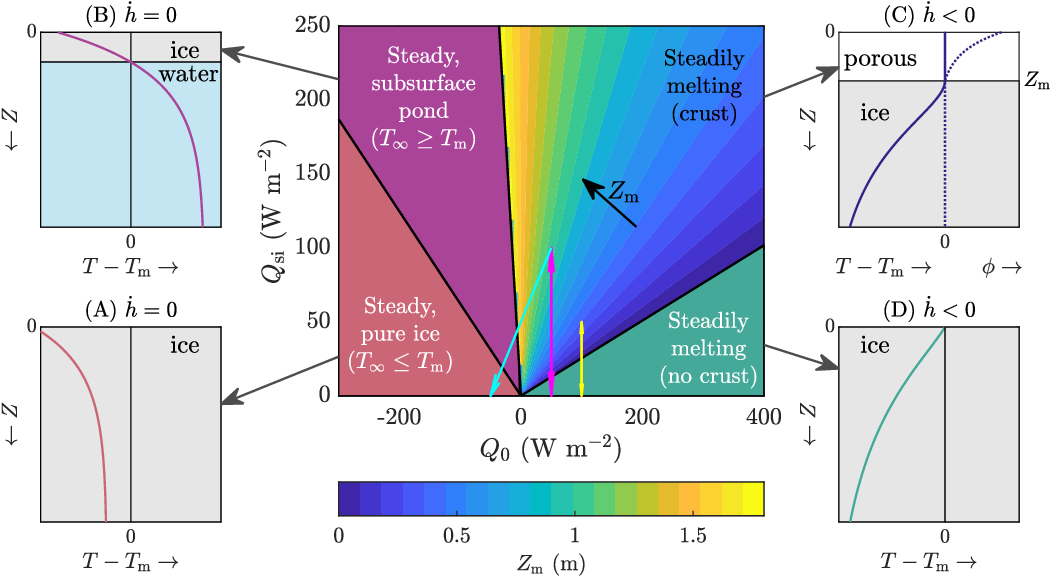}
    \caption{\scriptsize The types of steady ($\dot{h}=0$) and steadily melting ($\dot{h}<0$) solutions that are possible for different values of $Q_\mathrm{si}$ and $Q_0$. The plot shows the regions of the $Q_\mathrm{si}$-$Q_0$ plane for which there are: (A) steady (no melt) solutions with pure ice, (B) steady (no melt) solutions with a subsurface pond, (C) steadily melting solutions with a porous weathering crust, and (D) steadily melting solutions with no crust. Contours of weathering crust thickness $Z_\mathrm{m}$ are shown in region C. Typical temperature profiles in each region are shown, with the porosity in the weathering crust also shown in region C. In regions A and B, the far-field temperature $T_\infty$ is determined as part of the solution. In regions C and D, $T_\infty = -10^\mathrm{o}$C is prescribed. All other parameter values are as listed in Appendix \ref{sec:parameters_app}. The coloured arrows in the main plot show the change in forcing used in Figures \ref{fig:growth}, \ref{fig:removal_melt}, \ref{fig:periodic_daily_annual_compare} (magenta), Figure \ref{fig:removal_freezing} (cyan) and Figure \ref{fig:periodic_removal} (yellow). Adapted from \citet{Woods2023}.}
    \label{fig:Qsi_Q0_regions}
\end{figure}

These ideas about the conditions under which a weathering crust forms or not were demonstrated by \citet{Woods2023} using steadily melting solutions to the one-dimensional weathering crust problem. These are solutions to the model in Section \ref{sec:model} with constant forcings $Q_\mathrm{si}$ and $Q_0$ such that the surface lowers at a constant rate $\dot{h} < 0$, the weathering crust occupies a region between the surface and a constant depth $Z_\mathrm{m}$, and the temperature and porosity profiles in the $Z$-frame are constant in time. The ice melts in such a way that it appears the column of ice is simply being translated downwards over time (see Figure 2 in \citet{Woods2023}). The bottom of the weathering crust is located at $z=z_\mathrm{m} = h - Z_\mathrm{m}$. Figure \ref{fig:Qsi_Q0_regions} \citep[modified from][]{Woods2023} shows the region of $Q_0$-$Q_\mathrm{si}$ parameter space for which steadily melting solutions with a weathering crust exist (region C) and the corresponding weathering crust depth. It is also possible to have steadily melting states with no weathering crust, just solid ice (region D). Equations for the boundaries of region C are presented by \citet{Woods2023}.

\subsection{Steady solutions (non-melting)}
In regions A and B of Figure \ref{fig:Qsi_Q0_regions}, there is not enough energy to melt the surface, so a steadily melting state is not possible. Instead, it is possible to have steady states with no surface lowering ($\dot{h}=0$). Depending on the values of $Q_\mathrm{si}$ and $Q_0$, these steady states consist of either: pure ice (region A), or a layer of surface ice below which there is pure water (region B). The latter is not physically realistic, but is included here for completeness.

The temperature profile in the non-melting steady states is found by solving the energy equation \eqref{eq:temp_eqn} in the steady state. This gives
\begin{equation}
    T = T_\mathrm{m} + \frac{1}{\upsilon} \big( (1-a)Q_\mathrm{si} + Q_0 \big) +\frac{1}{\kappa k} (1-\chi)(1-a) Q_\mathrm{si} \big( 1 - e^{-\kappa Z} \big).
\end{equation}
In the steadily melting solutions, ice is being advected towards the surface in the $Z$-frame as a result of the physical surface $z=h(t)$ lowering at rate $-\dot{h}>0$. This allows us to prescribe the temperature $T_\infty$ at depth. In contrast, there is no advection in the non-melting steady states, so it is not possible to prescribe the temperature at depth, but only to prescribe that $\frac{\partial T}{\partial Z}$ tends to zero as $Z \to \infty$. The far-field temperature $T_\infty$ is instead determined as part of the solution. The sign of $T_\infty - T_\mathrm{m}$ determines whether we have a pure-ice solution ($T_\infty \leq T_\mathrm{m}$, region (b)) or subsurface pond solutions ($T_\infty \geq T_\mathrm{m}$), with the boundary between the two regions given by the line
\begin{equation}
    Q_0 + \bigg( \frac{\upsilon}{k \kappa} (1-\chi) + 1\bigg) (1-a) Q_\mathrm{si}=0.
\end{equation}

\subsection{Time-dependent solutions}
The key result from above is that it is the balance between shortwave radiation $Q_\mathrm{si}$ and longwave radiation and turbulent heat fluxes $Q_0$ that controls the growth and decay of the weathering crust. However, real ice sheets rarely have constant forcing. In the following sections, we use the time-dependent model to explore the dynamic growth and decay of the weathering crust when conditions transition between those favouring growth and those favouring decay.

In the time-dependent solutions we are seeking, we remove the assumption made in the steadily melting solutions that the surface of the ice is melting. It is possible to have a solid layer of ice above the top of the weathering crust (see Figure \ref{fig:transient_config_schematic}(b)). In this case, we call the location of the top of the weathering crust (the porous region) $z=z_\mathrm{f} = h - Z_\mathrm{f}$, where $Z_\mathrm{f}$ is the depth of the top of the weathering crust below the surface. The weathering crust thickness is defined as $Z_\mathrm{t} = Z_\mathrm{m} - Z_\mathrm{f}$. Schematics of possible transient configurations are shown in Figure \ref{fig:transient_config_schematic}.

\color{black}

\section{`Switching' forcing - growth and decay of the weathering crust}
\label{sec:switching}

\subsection{Setup}
To gain some initial understanding of the growth and decay of the weathering crust, we consider `switching' experiments, in which we start with a steadily melting state (with constant forcings $Q_\mathrm{si}^\mathrm{old}$, $Q_0^\mathrm{old}$), then instantaneously switch to a new set of constant forcings ($Q_\mathrm{si}^\mathrm{new}$, $Q_0^\mathrm{new}$), and investigate how the weathering crust responds. Such changes in forcing are demonstrated in Figures \ref{fig:growth}, \ref{fig:removal_melt}, \ref{fig:removal_freezing}.

Based on Figure \ref{fig:Qsi_Q0_regions}, we expect that a steadily melting weathering crust can be removed by (i) melting from the surface if we switch to forcings in region D (low $Q_\mathrm{si}$, positive $Q_0$), and by (ii) freezing from the surface if we switch to forcings in regions A or B (negative enough $Q_0$).

To grow a weathering crust, we can do the reverse: switch to forcings for which there is a steadily melting crust solution.

\subsection{Growth of a weathering crust}

\begin{figure}
    \centering
    \includegraphics[width=\textwidth]{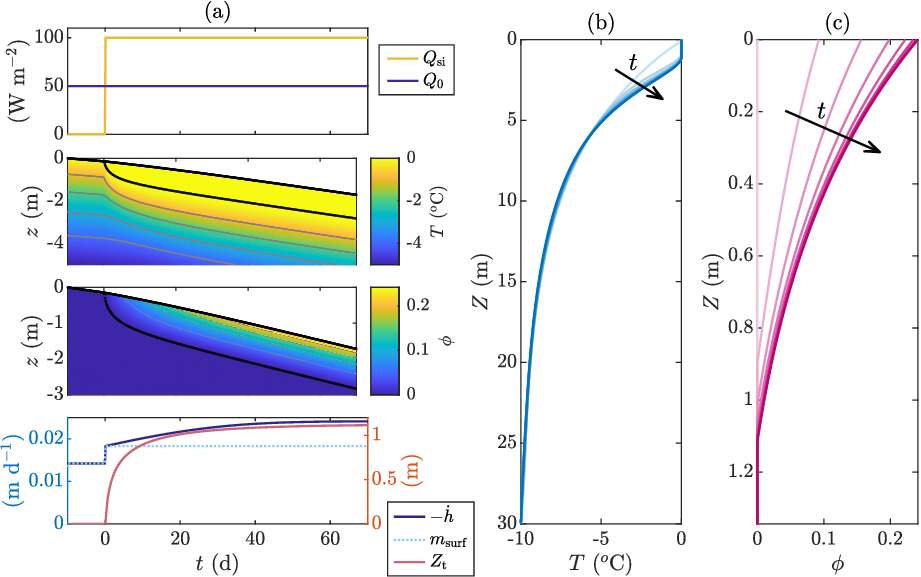}
    \caption{\scriptsize Growth of a weathering crust from a steadily melting, no-crust solution. The second and third plots in panel (a) show the evolution of the temperature $T$ and porosity $\phi$ profiles over times. The bottom of the weathering crust, $z=z_\mathrm{m}$, is shown as a black line. Contours of $T$ ($1^\mathrm{o}$C spacing) and $\phi$ (0.05 spacing) are shown in grey. The corresponding forcings $Q_\mathrm{si}$, $Q_0$, and the resulting surface lowering $-\dot{h}$, surface melt rate $m_\mathrm{surf}$ and the weathering crust thickness $Z_\mathrm{t}$ are shown in the first and fourth plots. Snapshots are shown of (b) temperature and (c) porosity at 10 day intervals, from day 0 to day 70. The forcings were switched at time $t=0$.}
    \label{fig:growth}
\end{figure}

We first consider the growth of a weathering crust from a no-crust steadily melting state, as shown in Figure \ref{fig:growth}. We start with no shortwave radiation $Q_\mathrm{si}$, then turn it on to $Q_\mathrm{si}=100$ W m$^{-2}$ whilst keeping $Q_0=50$ W m$^{-2}$ fixed. The snapshots of the temperature and porosity profiles in panels (b) and (c) show that the weathering crust develops to close to its final thickness quite quickly (over a few days), but with very low porosity initially. The porosity in the crust then continues to increase, as does the temperature in the top $\sim 6$ m of the ice, with the crust thickness changing very little. The resulting porosity and temperature profiles are always roughly exponential, due to the exponential decay of the internal shortwave radiation $F$ \eqref{eq:F}, which acts as the internal heat source.

\subsection{Removal via surface melting}
\label{sec:removal_melting}

\begin{figure}
    \centering
    \includegraphics[width=\textwidth]{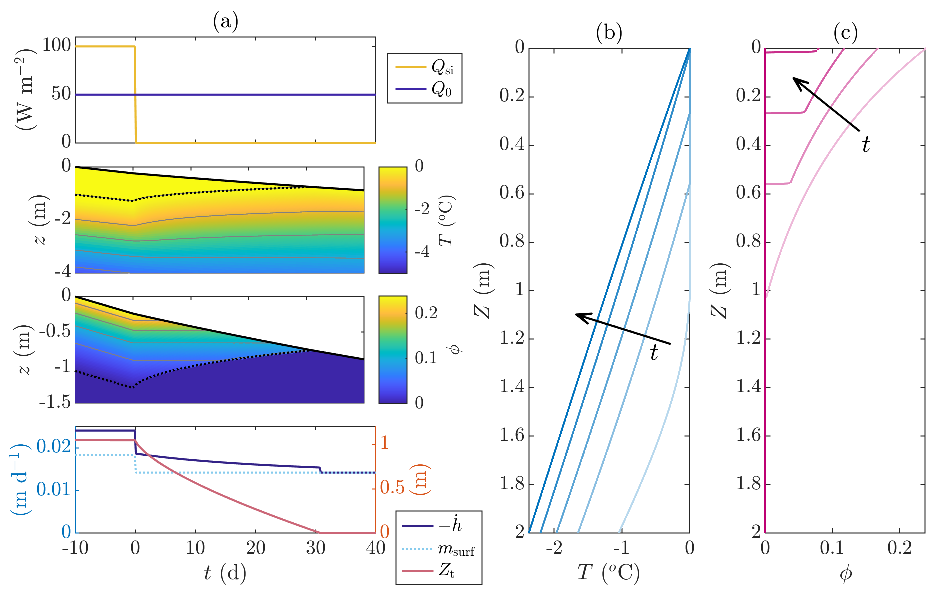}
    \caption{\scriptsize 
    Removal of a steadily melting weathering crust by melting from the surface. The second and third plots in panel (a) show the evolution of the temperature $T$ and porosity $\phi$ profiles over times. The bottom of the weathering crust, $z=z_\mathrm{m}$, is shown as a black line. Contours of $T$ ($1^\mathrm{o}$C spacing) and $\phi$ (0.05 spacing) are shown in grey. The corresponding forcings $Q_\mathrm{si}$, $Q_0$, and the resulting surface lowering $-\dot{h}$, surface melt rate $m_\mathrm{surf}$ and the weathering crust thickness $Z_\mathrm{t}$ are shown in the first and fourth plots. Snapshots are shown of temperature (b) and porosity (c) at 10 day intervals, from day 0 to day 40. The forcings were switched at time $t=0$.}
    \label{fig:removal_melt}
\end{figure}

We now consider removing the weathering crust by melting from the surface. Figure \ref{fig:removal_melt} shows the reverse experiment to that shown in Figure \ref{fig:growth}. This time, we are removing a steadily melting crust by turning off the shortwave radiation $Q_\mathrm{si}$ - an exact reversal of the forcings just considered. We see that removal happens through a combination of melting from the surface and freezing from below. Also note, from the porosity contour plot, that the porosity at each point in space remains unchanged until that point is either frozen or melted. That is, the lines of constant $\phi$ lie horizontal in the physical $z$ frame. The only way the porosity can change in the bulk of the domain is through melting due to the internal heat source provided by the internal radiation $F$. Therefore, when $Q_\mathrm{si}$ (and hence $F$) is turned off, the internal melting ceases. Freezing also changes the porosity, but this can only happen at interfaces between porous and solid ice (e.g. at $Z=Z_\mathrm{m}$), and results in a jump in porosity (see Figure \ref{fig:removal_melt}(c)). The jump in porosity satisfies the Stefan condition (in the $z$ frame)
\begin{equation}
\label{eq:Stefan_cond}
-k \frac{\partial T}{\partial z} = - \rho \mathcal{L} \phi \dot{z}_\mathrm{m} \quad \mathrm{at} \quad z=z_\mathrm{m},
\end{equation}
where $z_\mathrm{m} = h - Z_\mathrm{m}$ is the bottom of the weathering crust, $\phi$ is the porosity value on the crust side of the interface (with $\phi=0$ on the other side), and the temperature derivative is that on the no-crust side (being equal to zero on the crust side). The Stefan condition \eqref{eq:Stefan_cond} is satisfied automatically by the solution of the enthalpy equation when discretised in a conservative way.

The slight jump in the surface lowering rate $-\dot{h}$ that can be observed at the instant the crust is removed (around $t=30$ d) can be explained by a combination of (1) the freezing from below causing a discontinuity of $\phi$ at the freezing front at the bottom of the crust, and (2) the surface lowering rate $-\dot{h}$ depending on the surface porosity $\phi$ (with $m_\mathrm{surf}$ being constant). When the final thickness of crust is removed, the surface porosity jumps from non-zero to zero, but $m_\mathrm{surf}$ remains the same, resulting in a jump in $\dot{h}$. Physically, this is because the same amount of energy is now being used to melt solid ice rather than porous ice, so a smaller decrease in surface height is achieved for the same amount of energy (see discussion).

\subsection{Removal via surface freezing}

\begin{figure}
    \centering
    \includegraphics[width=\textwidth]{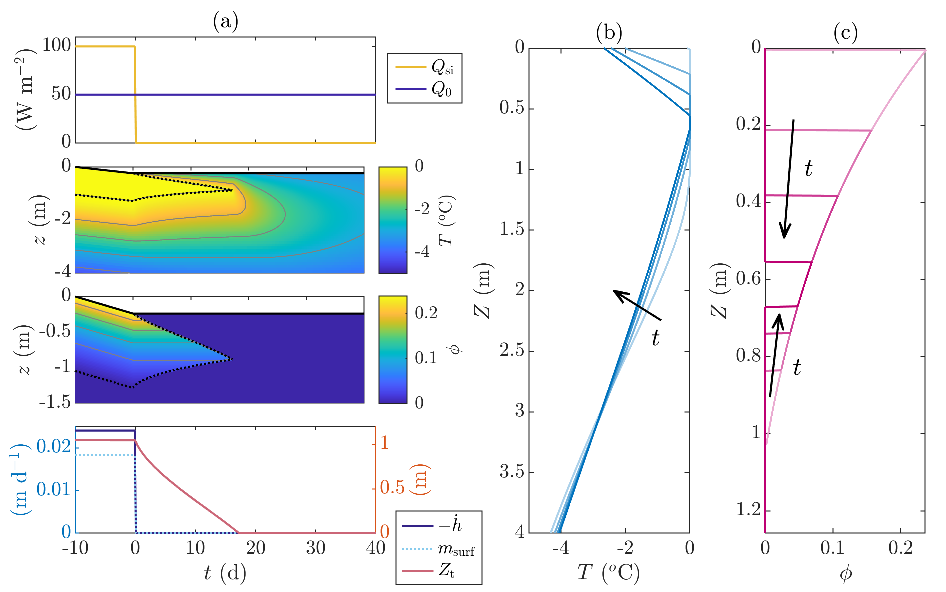}
    \caption{\scriptsize Removal of a steadily melting weathering crust by melting from the surface. The second and third plots in panel (a) show the evolution of the temperature $T$ and porosity $\phi$ profiles over times. The bottom of the weathering crust, $z=z_\mathrm{m}$, is shown as a black line. Contours of $T$ ($1^\mathrm{o}$C spacing) and $\phi$ (0.05 spacing) are shown in grey. The corresponding forcings $Q_\mathrm{si}$, $Q_0$, and the resulting surface lowering $-\dot{h}$, surface melt rate $m_\mathrm{surf}$ and the weathering crust thickness $Z_\mathrm{t}$ are shown in the first and fourth plots. Snapshots are shown of temperature (b) and porosity (c) at 5 day intervals, from day 0 to day 15. The forcings were switched at time $t=0$.}
    \label{fig:removal_freezing}
\end{figure}

The final case we consider is the switching experiment in which we remove a steadily melting weathering crust by freezing from the surface, as well as from below. We start with the same steadily melting state as used in the removal-by-melting test, and again switch to $Q_\mathrm{si}=0$ W m$^{-2}$, but this time the new value of $Q_0$ takes negative, instead of positive, values - we are moving into the left-hand region of Figure \ref{fig:Qsi_Q0_regions} in which there is no surface melting. The solution is shown in Figure \ref{fig:removal_freezing}. In this case, the upper boundary of the weathering crust does not coincide with the ice surface $z=h(t)$. Instead, there is a layer of solid ice above the weathering crust, and a freezing front at $z=z_\mathrm{f}(t)$. As in Figure \ref{fig:removal_freezing}, when $Q_\mathrm{si}=0$ W m$^{-2}$, there is no internal melting, so the porosity at each point in the crust remains unchanged until it jumps to zero when it is passed by one of the freezing fronts (the top $z=z_\mathrm{f}$, or the bottom $z=z_\mathrm{m}$). As a result, the porosity profile simply gets `truncated' over time, as shown particularly in Figure \ref{fig:removal_freezing}(c).

Despite the internal heat source being removed (by turning off the shortwave radiation $Q_\mathrm{si}$), the temperature deeper in the ice increases over time (see Figure \ref{fig:removal_freezing}(b)). This is because the more dominant effect is that the surface lowering has stopped, meaning that there is no longer any advection of cold ice at depth upwards towards the surface. The freezing of the water in the weathering crust releases heat, which conducts downwards and initially warms the ice lower down. Meanwhile the cooling of the surface due to negative $Q_0$ causes the temperature near the surface to decrease.

\subsection{Dependence on the value of $Q_0^\mathrm{new}$}

\begin{figure}
    \centering
    \includegraphics[width=\textwidth]{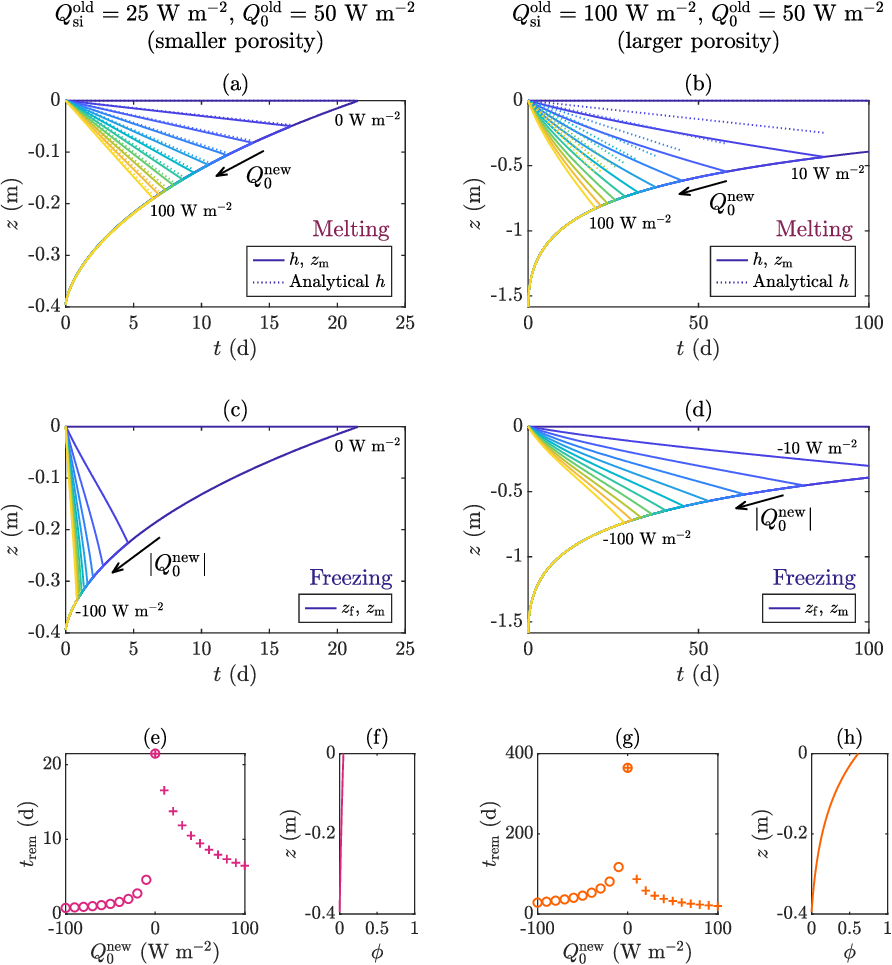}
    \caption{\scriptsize Removal of the weathering crust by ((a), (b)) melting from the surface and ((c), (d)) freezing from the surface, for different values of $Q_0$. Initially, there was a steadily melting crust with $Q_\mathrm{si}=25$ W m$^{-2}$ and $Q_0 = 50$ W m$^{-2}$ ((a), (c), (e), (f)), and $Q_\mathrm{si}=100$ W m$^{-2}$ and $Q_0 = 50$ W m$^{-2}$ ((b), (d), (g), (h)). In both cases, at $t=0$, the forcings were switched to $Q_\mathrm{si}^\mathrm{new}=0$ W m$^{-2}$ and values of $|Q_0^\mathrm{new}|$ from 0 to 100 W m$^{-2}$, in steps of 10 W m$^{-2}$ (positive for melting, negative for freezing). Solid lines show the top ($z=h(t)$ when melting, $z=z_\mathrm{f}$ when freezing) and bottom ($z=z_\mathrm{m}$) of the weathering crust for different values of $Q_0$. The analytical approximation \eqref{eq:hdot_ana_small_phi} for $z=h(t)$ for small $\phi$ is also shown (dotted). Panels (e) and (g) show how the time $t_\mathrm{rem}$ taken to remove the crust depends on $Q_0^\mathrm{new}$. Panels (f) and (h) shown the porosity profiles of the initial steadily melting solutions, taking values of 0.05 and 0.61 at the surface, respectively.}
    \label{fig:vary_Q0}
\end{figure}

We also explore how the \emph{rate} of the removal of a weathering crust depends on the change in the values of $Q_\mathrm{si}$ and $Q_0$. Firstly, we investigate the impact of $Q_0^\mathrm{new}$. We explore what happens when we turn off the shortwave radiation (set $Q_\mathrm{si}^\mathrm{new}=0$ W m$^{-2}$) and vary the new value of $Q_0^\mathrm{new}$ (Figure \ref{fig:vary_Q0}). We consider both melting ($Q_0^\mathrm{new}>0$) and freezing ($Q_0^\mathrm{new}<0$) from the surface. (Figures \ref{fig:removal_melt} and \ref{fig:removal_freezing} have already shown examples of such solutions for the specific cases $Q_0^{\mathrm{new}} = 50$ W m$^{-2}$ and $Q_0^{\mathrm{new}} = -50$ W m$^{-2}$.)

Broadly, we see that the time taken to remove the weathering crust by melting is shorter when $Q_0^\mathrm{new}$ is more positive (`+' in Figure \ref{fig:vary_Q0} (e), (g)), and the time taken to remove the weathering crust by freezing is shorter when $Q_0^\mathrm{new}$ is more negative (`o' in Figure \ref{fig:vary_Q0} (e), (g)).

Comparing the melting and freezing cases, we see that the freezing from the surface removes the weathering crust much more quickly than melting from the surface when the porosities are small ($\leq0.05$, left-hand side of Figure \ref{fig:vary_Q0}). Smaller porosities mean that much more energy is required to melt the ice fraction (0.95) than freeze the water fraction (0.05). On the contrary, when the porosities are larger (0.61 at the surface initially, right-hand side of Figure \ref{fig:vary_Q0}), the removal times are more similar. This time, melting removes the crust more quickly than freezing, since there is less ice to melt than water to freeze.

The fact that it takes less time to remove the weathering crust by freezing from the surface when $Q_0^\mathrm{new}$ is more negative can be explained by the surface energy balance \eqref{eq:SEB_simplified} (with $m_\mathrm{surf}=0$ in this case). To maintain the surface energy balance, a decrease in $Q_0$ must be compensated for by either decreasing the surface temperature $T$ or by increasing the heat flux $k \frac{\partial T}{\partial Z}$ towards the surface from the ice below. This acts to cool the internal ice.

The reason that the time taken to remove the crust by melting is shorter when $Q_0^\mathrm{new}$ is more positive is that increasing $Q_0$ increases the amount of surface melting. The surface melt rate given by the surface energy balance \eqref{eq:SEB_simplified} (when $Q_\mathrm{si}=0$, as here) is
\begin{equation}
\label{eq:msurf}
m_\mathrm{surf} = \frac{Q_0}{\rho \mathcal{L}}.
\end{equation}
This in turn leads to an increase in the rate of surface lowering $- \dot{h}$, according to the kinematic condition \eqref{eq:kinematic_ice}. For small surface porosity, the surface lowering rate can be approximated analytically from \eqref{eq:kinematic_ice} as
\begin{equation}
\label{eq:hdot_ana_small_phi}
-\dot{h} = \frac{m_\mathrm{surf}}{1- \phi} \approx m_\mathrm{surf},
\end{equation}
giving an approximately constant rate of surface lowering. This is observed in Figure \ref{fig:vary_Q0}(a), where there is good agreement between the analytical (dotted) and numerical (solid) solutions for $z=h$. In this case, the small $\phi$ assumption is valid, with $\phi \approx 0.05$ at the surface in the initial steadily melting state. However, the agreement is less good in Figure \ref{fig:vary_Q0}(b) where $\phi \approx 0.61$ at the surface, making the small $\phi$ assumption less valid.

We also observe that the evolution of the bottom of the weathering crust $z=z_\mathrm{m}$ is independent of the value of $Q_0$. We can see this mathematically by writing down the problem for the temperature in the solid ice below the weathering crust in the absence of shortwave radiation. This is
\begin{equation}
\label{eq:below_crust_heat_eqn}
\rho c \frac{\partial T}{\partial t} = k \frac{\partial^2 T}{\partial z^2}
\end{equation}
in $z<z_\mathrm{m}(t)$, with three boundary conditions: the far-field condition \eqref{eq:Tinf_BC} as $z \to -\infty$, and the Stefan condition \eqref{eq:Stefan_cond}, together with continuity of temperature $T = T_\mathrm{m}$, at $z = z_\mathrm{m}$.

We have one more boundary condition than required to solve the heat equation, enabling us to also solve for the moving boundary $z=z_\mathrm{m}(t)$. We note that this problem is independent of the boundary conditions at $z=h$, hence $z=z_\mathrm{m}$ can be solved for independent of $Q_0$ (although, with the given initial temperature and porosity profiles, it is unfortunately not straightforward to solve analytically).

\subsection{Dependence on the value of $Q_\mathrm{si}^\mathrm{new}$}
\begin{figure}
    \centering
    \includegraphics[width=\textwidth]{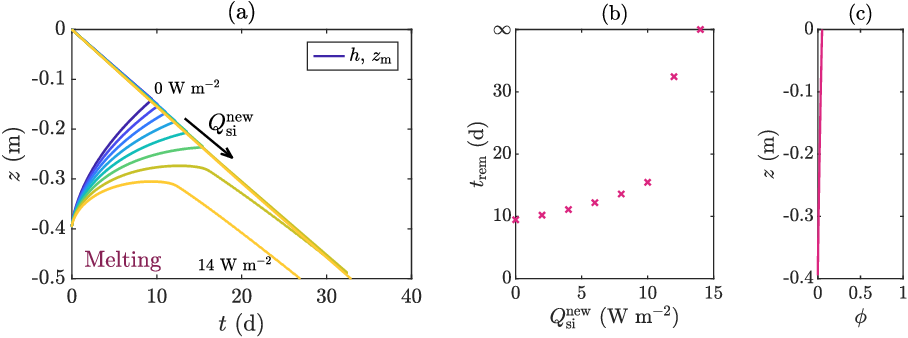}
    \caption{\scriptsize Removal of the weathering crust by melting from the surface for different values of $Q_\mathrm{si}$. Initially, there was a steadily melting crust with $Q_\mathrm{si}=25$ W m$^{-2}$ and $Q_0 = 50$ W m$^{-2}$. The forcings were switched at $t=0$ to $Q_0 = 50$ W m$^{-2}$ (unchanged) and various values of $Q_\mathrm{si}$, from 0 to 14 W m$^{-2}$, in steps of 2 W m$^{-2}$. Panel (a) shows the evolution of the top (solid, $z=h$) and bottom (dashed, $z=z_\mathrm{m}$) of the weathering crust for different values of $Q_\mathrm{si}^{\mathrm{new}}$. Panel (b) show how the time $t_\mathrm{rem}$ taken to remove the crust depends on the new value of $Q_\mathrm{si}$. Panel (c) shows the porosity profile of the initial steadily melting solutions, taking a value of 0.05 at the surface.}
    \label{fig:vary_Qsi}
\end{figure}

We now explore how the removal of the weathering crust is affected by the new value of $Q_\mathrm{si}$, which affects both the internal melting (via the internal radiation $F$ \eqref{eq:F}) and surface melting (via the surface energy balance \eqref{eq:SEB_simplified}). We do this by starting with the steadily melting state with $Q_\mathrm{si}^\mathrm{old}=25$ W m$^{-2}$ and $Q_0^\mathrm{old} = 50$ W m$^{-2}$, then switching $Q_\mathrm{si}$ to a new, lower value, whilst keeping $Q_0$ fixed. Figure \ref{fig:vary_Qsi}(a) shows that the effect of this is to change the rate of freezing at $z=z_\mathrm{m}$, whereas the rate of surface lowering changes very little. The surface lowering is approximately constant in this example because all the values of $Q_\mathrm{si}^\mathrm{new}$ are small (ranging from 0 to 14 W m$^{-2}$), so the surface melt rate \eqref{eq:msurf} is dominated by the $Q_0$ term. This is amplified by $Q_\mathrm{si}$ being multiplied by $\chi (1-a) \approx 0.2 < 1$, representing the fact that not all the shortwave radiation is absorbed at the surface: some is reflected ($a Q_\mathrm{si}$) and some is absorbed internally ($(1-\chi)(1-a)Q_\mathrm{si}$). If $Q_0^\mathrm{new}$ was closer to 0 and $Q_\mathrm{si}^\mathrm{new}$ was larger, the change in surface lowering with $Q_\mathrm{si}^\mathrm{new}$ would be more noticeable.

The main effect of varying $Q_\mathrm{si}^\mathrm{new}$ is to change the internal heat source, generated by the internal shortwave radiation $F$, given by \eqref{eq:F}. The smaller  $Q_\mathrm{si}^\mathrm{new}$ is, the greater is the reduction in the internal heat source (as $Q_\mathrm{si}$ switches from $Q_\mathrm{si}^\mathrm{old}$ to $Q_\mathrm{si}^\mathrm{new}$ in \eqref{eq:F}), and hence the crust gets frozen from below more quickly. There is a maximum value of  $Q_\mathrm{si}^\mathrm{new}$ for which the crust gets removed (12.7 W m$^{-2}$ here), given by the boundary between regions C and D in the contour plot in Figure \ref{fig:Qsi_Q0_regions} (with the equation for this line given by \citet{Woods2023}). Any value larger than this results in a new steadily melting crust of a different thickness being formed, determined by Figure \ref{fig:Qsi_Q0_regions}. As $Q_\mathrm{si}^\mathrm{new}$ tends towards its maximum value for removal, the removal time tends to infinity (Figure \ref{fig:vary_Qsi}(b)).

\section{Periodic forcing}
\label{sec:periodic}
Although the switching experiments in Section \ref{sec:switching} provide insight into the basic behaviours of weathering crust growth and removal, such sudden forcing transitions are not very realistic. In reality, we do not expect the forcings to remain constant for sustained periods of time. As previously discussed, the weathering crust and the weather conditions that force it are very variable, changing on the order of hours and days \citep[e.g.][]{Fausto2021}, with the weathering crust growing and decaying over short timescales as a result \citep{Schuster2001}. Therefore, a more relevant test is sinusoidal variations of forcings, representing, for example, diurnal or annual variations.

In all of the experiments considered in this section, a sinusoidal forcing is run until a roughly periodic state is reached. Note that the forcings here are not necessarily realistic diurnal or annual variations (since these are not sinusoidal in amplitude), but serve to demonstrate the kind of periodic behaviour that is possible from our model. As noted by \citet{Schuster2001}, the initial conditions of the weathering crust (e.g. how porous it is) will likely have a strong effect on the amount and type (surface or internal) of melt resulting from given forcing conditions. Periodic solutions provide a way to examine the behaviour of the weathering crust whilst eliminating the impact of the initial conditions.

\subsection{Periodic growth and complete removal}

\begin{figure}
    \centering
    \includegraphics[width=\textwidth]{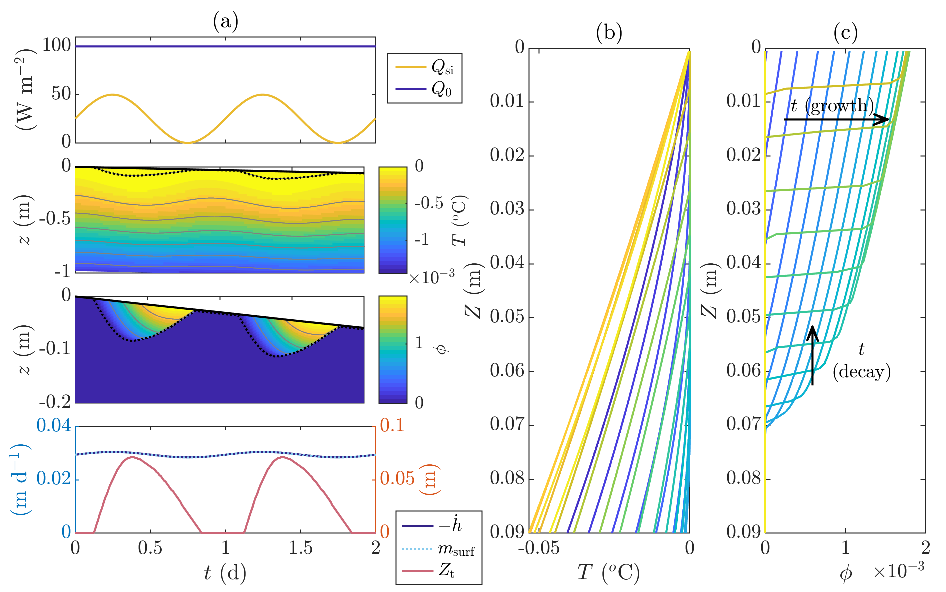}
    \caption{\scriptsize Periodic weathering crust generated by sinusoidal forcing, with complete removal each period by surface melting. The second and third plots of panel (a) show the evolution of the temperature $T$ and porosity $\phi$ profiles over times. The bottom of the weathering crust, $z=z_\mathrm{m}$, is shown by a black line. Contours of $T$ ($0.2^\mathrm{o}$C spacing) and $\phi$ ($0.5 \times 10^{-3}$ spacing) are shown in grey. The corresponding forcings $Q_\mathrm{si}$, $Q_0$, and the resulting surface lowering $-\dot{h}$, surface melt rate $m_\mathrm{surf}$ and weathering crust thickness $Z_\mathrm{t}$ are shown in the first and fourth plots. Snapshots are shown of temperature (b) and porosity (c) at 1 hour intervals over the course of 1 day.}
    \label{fig:periodic_removal}
\end{figure}

Figure \ref{fig:periodic_removal} shows an example where the weathering crust is periodically completely removed (by melting from the surface and freezing from below), then regrown, as a result of daily variations in $Q_\mathrm{si}$, with $Q_0$ remaining constant. In this example, the porosity in the weathering crust is very small ($\sim 10^{-3}$), due to the limited time for it to grow. A key point to note is the non-symmetric behaviour in the growth and removal of the crust, with the crust growing at a faster rate than it decays. We see that when the crust is growing (blue to green snapshots in panel (c)), the porosity smoothly increases from zero at the bottom of the crust, with the depth of the crust and the porosity at each depth increasing smoothly (as in Figure \ref{fig:growth}). In contrast, when the crust is removed, the freezing from below results in a jump in the porosity at the bottom of crust, according to the Stefan condition \eqref{eq:Stefan_cond} (as in Figures \ref{fig:removal_melt} and \ref{fig:removal_freezing}). Compared to the switching examples in Figures \ref{fig:removal_melt} and \ref{fig:removal_freezing} in which the $Q_\mathrm{si}=0$, here $Q_\mathrm{si}>0$, so there continues to be internal melting whilst freezing from the bottom of the crust is occurring. We see this in the fact that the lines of constant porosity slope downwards in the $t$-$z$ plane here, compared to being horizontal in Figures \ref{fig:removal_melt} and \ref{fig:removal_freezing}. 
Also note the lag in the crust thickness. The crust thickness continues to increase for a short while after $Q_\mathrm{si}$ has begun to decrease, and $Q_\mathrm{si}$ almost reaches its maximum before the crust begins to re-form. This is explained by the fact that the weathering crust does not respond instantaneously to the changes in $Q_\mathrm{si}$ and $Q_0$, as demonstrated in the switching experiments in Section \ref{sec:switching}. It takes time for heat to conduct through the domain.

\subsection{Diurnal vs. annual forcing}

\begin{figure}
    \centering
    \includegraphics[width=\textwidth]{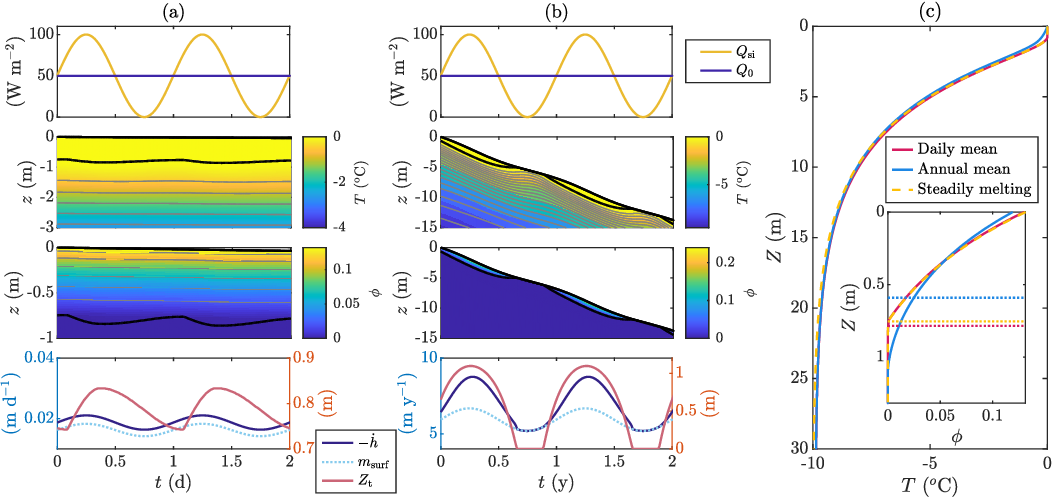}
    \caption{\scriptsize Periodic weathering crust generated by periodic forcing, for the same forcings but over different timescales (a) diurnal and (b) annual. The simulations have been run to reach a roughly periodic state: 100 days for the daily forcings, 10 years for the yearly. The second and third plots in panels (a) and (b) show the evolution of the temperature $T$ and porosity $\phi$ profiles over times. The bottom of the weathering crust, $z=z_\mathrm{m}$, is shown by a black line. Contours of $T$ ($0.5^\mathrm{o}$C spacing) and $\phi$ ($0.02$ spacing) are shown in grey. The corresponding forcings $Q_\mathrm{si}$, $Q_0$, and the resulting surface lowering $-\dot{h}$, surface melt rate $m_\mathrm{surf}$ and weathering crust thickness $Z_\mathrm{t}$ are shown in the first and fourth plots. Panel (c) compares the time mean solutions to the steadily melting solution for the mean forcings.}
    \label{fig:periodic_daily_annual_compare}
\end{figure}

We now explore the impact of the period of the forcing. Figure \ref{fig:periodic_daily_annual_compare} shows two periodic solutions forced by the same periodic forcings ($Q_\mathrm{si}$ varying between 0 and 100 W m$^{-2}$, $Q_0$ constant at 50 W m$^{-2}$), but with different periods: 1 day and 1 year. We see that with the 1 day period, the crust thickness undergoes small variations ($\sim 0.1$ m), with the crust remaining present and of approximately constant thickness throughout. With the 1 year period, the crust gets completely removed (and regrown) periodically. This is in line with the switching experiment in Figure \ref{fig:removal_melt}, in which it took $\sim 1$ month to remove the steadily melting weathering crust corresponding to $Q_\mathrm{si}=100$ W m$^{-2}$, $Q_0 = 50$ W m$^{-2}$ by turning off the shortwave radiation (i.e. the switching version of the periodic forcing in Figure \ref{fig:periodic_daily_annual_compare}).

We can also compare the periodic solutions to steadily melting solutions for the same mean forcing (Figure \ref{fig:periodic_daily_annual_compare}(c)). We see that the mean $\phi$, $T$ and $Z_\mathrm{m}$ of the periodic solution are close to, but not equal to, the steadily melting solution of the mean forcing ($Q_\mathrm{si}=50$ W m$^{-2}$, $Q_0=50$ W m$^{-2}$). Firstly, the mean porosity is non-zero as far down as the greatest depth to which the porosity is non-zero at any time during the period. This is deeper than the mean crust depth in the steadily melting porosity profile. The depth to which non-zero porosity is reached is largest for the annual forcing since there is more time for heat to conduct deeper into the ice. However, the annual mean crust thickness is lower than both the steadily melting crust thickness and the mean thickness for the diurnal period because the crust is completely removed for periods of time. The crust would continue to thin further during some of this time if there was more ice available to melt (i.e. if the crust was thicker in the first place). Since this is not possible, it leads to a proportionately thinner mean crust thickness. On the other hand, the mean crust thickness for the diurnal period is quite close to the steadily melting crust thickness. This is due to the diurnal variations being faster than the timescale over which the crust significantly evolves, so the oscillations in forcing are effectively averaged out. The mean crust thickness for the diurnal variations is slightly larger that the steadily melting crust thickness, and we attribute this to the asymmetric growth and decay of the weathering crust. As seen in Figure \ref{fig:periodic_removal}(a),(c), growth occurs more quickly than removal, favouring a crust that is thicker on average than if the constant mean of the forcing was applied.

\subsection{Superimposed diurnal and annual forcing}
\label{sec:diurnal_annual}
As well as considering diurnal and annual variations in isolation, it is informative to consider diurnal variations superimposed onto annual variations. This is representative of the increased solar radiation during the day compared to at night, and during the summer compared to winter. When modelling the evolution of a system that responds to the solar cycle, such as the weathering crust, we would like to know how important the short-term diurnal variations in forcing are. When considering longer timescales (many years, e.g. to investigate the response to changing climate conditions), it would be preferable to be able to ignore the diurnal variations, enabling us to take longer timesteps in numerical computations. Here, we explore the impact of these superimposed diurnal variations on the growth and decay of the weathering crust.

\begin{figure}
    \centering
    \begin{subfigure}{0.36\textwidth}
        \centering
        \includegraphics[width=\textwidth]{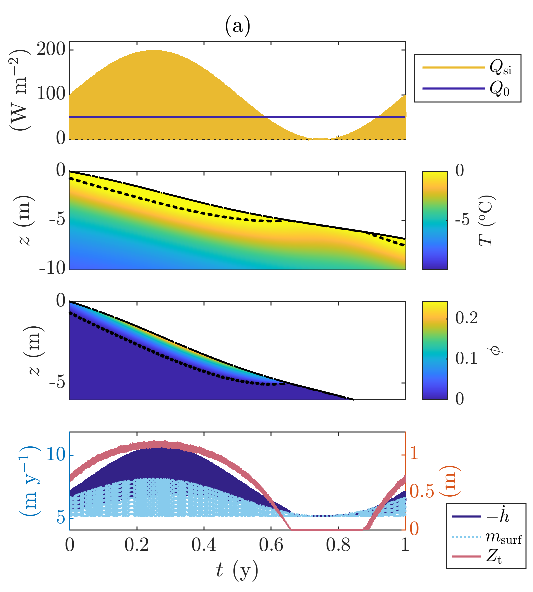}
    \end{subfigure}
    \hfill
    \begin{subfigure}{0.36\textwidth}
        \centering
        \includegraphics[width=\textwidth]{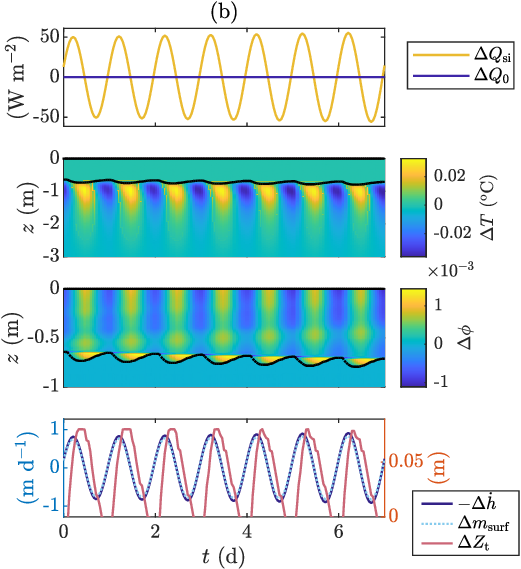}
    \end{subfigure}
    \hfill
    \begin{subfigure}{0.25\textwidth}
        \centering
        \includegraphics[width=\textwidth]{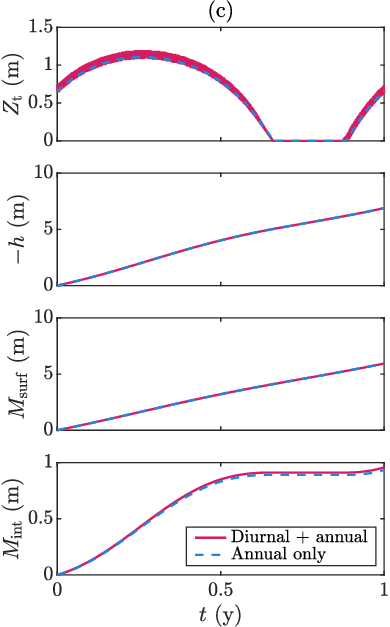}
    \end{subfigure}
    \caption{\scriptsize Periodic weathering crust generated by diurnal sinusoidal variations superimposed onto annual sinusoidal variations (panel (a)). The diurnal variations are such that $Q_\mathrm{si}$ goes to zero each night. The simulations have been run for 22 years to reach a periodic state. Panel (b) shows the \emph{difference} between the solution in panel (a) and the solution with only annual variations shown in Figure \ref{fig:periodic_daily_annual_compare} (`$\Delta=$ diurnal-and-annual -- annual-only'). The second and third plots in panels (a) and (b) show the evolution of the temperature $T$ and porosity $\phi$ profiles over time. The black dashed line shows the bottom of the weathering crust, $z=z_\mathrm{m}$. The corresponding forcings $Q_\mathrm{si}$, $Q_0$, and the resulting surface lowering $-\dot{h}$, surface melt rate $m_\mathrm{surf}$ and weathering crust thickness $Z_\mathrm{t}$ are shown in the first and fourth plots. Panel (c) shows weathering crust thickness $Z_\mathrm{t}$, the surface lowering $-h$, cumulative surface melt $M_\mathrm{surf}$, and cumulative internal melt $M_\mathrm{int}$ for the diurnal-superimposed-onto-annual forcings (panel (a)) and the purely annual forcings (Figure \ref{fig:periodic_daily_annual_compare}(b)).}
    \label{fig:superimposed_daily_annual_compare}
\end{figure}

To do this, we take the shortwave radiation to be
\begin{equation}
    Q_\mathrm{si}(t) = {Q}_\mathrm{si}^\mathrm{year}(t)(1 +  \sin (2 \pi t/t_\mathrm{day})),
\end{equation}
where ${Q}_\mathrm{si}^\mathrm{year}(t)$ is the annually-varying diurnal-mean,
\begin{equation}
    {Q}_\mathrm{si}^\mathrm{year}(t) = \bar{Q}_\mathrm{si} + A \sin(2 \pi t/t_\mathrm{year}),
\end{equation}
with mean $\bar{Q}_\mathrm{si}$ and amplitude $A$. Here, $t_\mathrm{day}$ is the period of 1 day, $t_\mathrm{year}$ is the period of 1 year. This form of $Q_\mathrm{si}$ takes a minimum of 0 in each daily cycle (at night). We take ${Q}_\mathrm{si}^\mathrm{year}(t)$ to be the same as the shortwave radiation forcing used for the annual variations in Figure \ref{fig:periodic_daily_annual_compare}, with $A=\bar{Q}_\mathrm{si}=50$ W m$^{-2}$. The result is shown in Figure \ref{fig:superimposed_daily_annual_compare}.

The main effect of superimposing diurnal variations onto the annual variations is to create diurnal oscillations in the porosity, temperature, weathering crust thickness, internal and surface melt rates, and surface lowering rate. The overall behaviour of the solution is similar to that with the annual variations only.

The oscillations in temperature are of order $0.01^\mathrm{o}$C, and the oscillations in porosity are of order $0.001$ (Figure \ref{fig:superimposed_daily_annual_compare}(b)). The oscillations in weathering crust thickness (Figure \ref{fig:superimposed_daily_annual_compare}(b)) are once again of order 0.1 m, consistent with the diurnal variations in Figure \ref{fig:periodic_daily_annual_compare}. As already seen in Figure \ref{fig:periodic_daily_annual_compare}(a),(c), the daily mean of these oscillations is slightly larger than the crust thickness variation resulting from the annual-only forcing (Figure \ref{fig:superimposed_daily_annual_compare}(c)), attributed to the asymmetric growth and decay of the weathering crust.

We see that the cumulative surface lowering $-h$ ($=\int -\dot{h} \ \mathrm{d} t$), cumulative surface melt $M_\mathrm{surf}$ ($=\int m_\mathrm{surf} \ \mathrm{d} t$) and cumulative internal melt $M_\mathrm{int}$ ($=\frac{1}{\rho} \int \int \max(m_\mathrm{int},0) \ \mathrm{d}z \ \mathrm{d}t$) over 1 year are largely unaffected by the diurnal variations (Figure \ref{fig:superimposed_daily_annual_compare}(c)).
The surface lowering has contributions from surface melting and internal melting. Internal melting lowers the porosity, and the resulting non-zero surface porosity means that more surface lowering occurs than would be accounted for by surface melting alone. Any already-liquid water at the surface (produced by internal melt) adds to the surface melt and runs off as the surface lowers. In general, not all the internal meltwater runs off at the surface; some of it can instead refreeze. In this example, however, the amount that refreezes is small, so almost all the internal meltwater is removed at the surface. Hence the surface lowering is approximately equal to the sum of the internal and surface melting. 

Nevertheless, slightly more internal melting does occur in the diurnal+annual solution than the annual-only solution, due to the diurnal refreezing and then re-melting at the bottom of the weathering crust. This is demonstrated by the difference in porosity between the solutions with and without diurnal variations (Figure \ref{fig:superimposed_daily_annual_compare}(b)). The amount of additional internal melting produced as a result of this diurnal freeze-melt cycle is small, since the porosity at the bottom of the weathering crust is close to zero. For larger variations in $Q_0$ as well as $Q_\mathrm{si}$, it can be larger.

\section{Conclusion and Discussion}
\label{sec:conclusion}
We have explored time-dependent solutions of the one-dimensional continuum model for the weathering crust we presented in \citet{Woods2023}. We have used an enthalpy method to numerically solve the model equations, removing the need to track the boundaries between the porous, pure-ice and pure-water regions of the domain. Our choice of idealised `switching' and sinusoidal forcings has enabled us to gain insight into the physical processes controlling the growth and decay of the weathering crust in response to changing weather conditions (represented in our model by the shortwave radiation $Q_\mathrm{si}$ and the longwave radiation and turbulent heat fluxes $Q_0$). We find that our model produces broadly realistic behaviours when compared qualitatively to observations (see below), suggesting that it could be used as a tool to make predictions about the response of the weathering crust to future weather and climate conditions. There are, however, improvements and extensions that could be made to the model, discussed below. 

\subsection{Growth and decay behaviour and timescales}
In line with observations \citep{Schuster2001,Muller1969}, we have seen that it is the balance between shortwave radiation ($Q_\mathrm{si}$) and longwave radiation and turbulent heat fluxes ($Q_0$) that controls the growth and decay of the weathering crust, with clear, sunny conditions (high $Q_\mathrm{si}$, low $Q_0$) favouring growth, and warm, windy, cloudy conditions favouring decay (low $Q_\mathrm{si}$, high $Q_0$). Furthermore, we have observed in our model different mechanisms of growth and decay. The weathering crust grows via internal melting (Figure \ref{fig:growth}), resulting from some shortwave radiation penetrating beneath the surface. Decay/removal of the crust can occur by (1) melting from the surface (Figure \ref{fig:removal_melt}), or (2) freezing from the surface (Figure \ref{fig:removal_freezing}), combined with (3) freezing from below (Figures \ref{fig:removal_melt} and \ref{fig:removal_freezing}).

Moreover, there is an asymmetry in the growth and decay behaviour, made clear by our study of sinusoidal forcings. Growth maintains a smooth porosity profile ($\phi \to 0$ smoothly at the bottom of the weathering crust), whereas decay by freezing (from above or below) results in a jump in porosity as phase change occurs. In a periodic setting, growth occurs more quickly than decay. This is true across a number of different cases: (1) when the weathering crust is periodically completely removed by surface melting (Figure \ref{fig:periodic_removal}), (2) when it oscillates in thickness but is never removed (Figure \ref{fig:periodic_daily_annual_compare}(a)), or (3) when it is periodically completely removed by surface freezing (not shown here). The asymmetry in timescales means that a weathering crust produced by a sinusoidal forcing about a mean has a thicker weathering crust on average than if the forcing was held constant at the mean value, unless the crust is completely removed during the cycle (Figure \ref{fig:periodic_daily_annual_compare}).

The timescale for removal in the specific examples studied here is typically on the order of several days. The removal time decreases with increasing magnitude of $Q_0$ (with $Q_0>0$ corresponding to removal by surface melting, and $Q_0<0$ corresponding to removal by surface freezing). The porosity, particularly at the surface, also affects the rate of removal of the weathering crust: a more porous crust is easier to remove by melting, and harder to remove by freezing.

\subsection{Actual vs. potential surface lowering}
In Section \ref{sec:removal_melting}, we observed that the surface porosity $\phi_\mathrm{surf}$ affects the rate of surface lowering, with a jump in $\phi$ at the surface resulting in a jump in the rate $-\dot{h}$ of surface lowering. In Section \ref{sec:diurnal_annual}, we saw that the surface lowering is less than the surface melting due to the non-zero surface porosity (see kinematic condition \eqref{eq:kinematic_ice}). This ties into ideas discussed by \citet{Schuster2001}. They studied the different stages of weathering crust growth and decay in terms of the ratio $R$ between the actual (observed) surface lowering and the potential surface lowering (based on the total energy available to melt). In our case,
\begin{equation}
R = \frac{\mathrm{actual \ surface \ lowering}}{\mathrm{potential \ surface \ lowering}} = \frac{-\dot{h}}{\big((1-a) Q_\mathrm{si} + Q_0 \big)/\rho \mathcal{L}},
\end{equation}
where the potential surface lowering comes from the total melt energy: the sum of the energy absorbed at the surface, $\chi (1-a) Q_\mathrm{si} + Q_0$, and the energy absorbed internally, $(1-\chi)(1-a)Q_\mathrm{si}$.
It is expected that $R<1$ during crust growth because some of the melt energy is used for internal melting, which is not visible at the surface. During crust decay (removal via melting), it is expected that $R>1$ because the porous ice requires less energy to remove than solid ice \citep{Schuster2001}. Phrasing this differently, the observed surface lowering would suggest more (less) melting is occuring during decay (growth) than is truly the case \citep{Muller1969}. This can lead to errors in surface ablation (mass loss) predictions based on observed surface lowering, although these errors resulting from short-term crust variations are expected to cancel out over longer timescales \citep{Muller1969}. 

Using the kinematic condition \eqref{eq:kinematic_ice} and the expression \eqref{eq:msurf} for the surface melt rate $m_\mathrm{surf}$, we can rewrite $R$ as
\begin{equation}
R = \Bigg(\frac{\chi (1-a) Q_\mathrm{si} + Q_0}{(1-a) Q_\mathrm{si} + Q_0}\Bigg) \Bigg( \frac{1}{1-\phi_\mathrm{surf}} \Bigg),
\end{equation}
where the first fraction is the proportion of the total energy available for melt that is used to melt the surface. This shows that the key factors causing a discrepancy between the observed and expected surface lowering are (1) the fact that some of the shortwave radiation penetrates and is absorbed below the surface, determined by the parameter $\chi$, and (2) the high surface porosity of the weathering crust.

\subsection{Future/ongoing work}
\subsubsection{Including drainage of internal meltwater}

One key process missing from the current model is lateral flow of meltwater \emph{within} the weathering crust, and the possibility for the crust to become partially saturated (have air as well as water in the pore space), as occurs in reality \citep{Cooper2018,Irvine-Fynn2021}. We assume that surface meltwater runs off instantaneously, but any internal meltwater can only stay in place and potentially refreeze later on. The modelling work of \citet{Hoffman2014} demonstrated that lateral drainage of internal meltwater in the weathering crust is vital for correctly reproducing observed surface lowering, porosity and temperature in an Antarctic setting. They found that, without lateral drainage, surface lowering and porosity are underestimated. The drainage of meltwater internally enables the weathering crust to become unsaturated and hence refreeze with non-zero porosity. Less energy is then required to melt the porous ice, compared to solid ice, during the next day/year. This results in increased surface lowering (see previous discussion of the effect of porosity on surface lowering).

In future work, we will explore this effect by including a parametrisation for lateral drainage in our one-dimensional model. Including a lateral drainage parametrsiation also requires including a model for gravity-driven vertical drainage, to fill the `gaps' left by the internally drained meltwater. The vertical and lateral drainage can both be modelled using Darcy's law for flow through a porous medium.

\subsubsection{Accumulation}
During the winter, snowfall will accumulate on the frozen surface on which the weathering crust will form in the summer. The layer of snow can insulate the ice below during the winter, and some energy that would otherwise be used to melt the ice is required to melt the snow at the start of the summer. Our model currently neglects such accumulation, but incorporating this is a target for future work.

\subsubsection{Real-world forcing}
The forcings used in this study were idealised `switching' and sinusoidal forcings. Although informed by the conditions under which we expect growth and removal of the weathering crust (see Figure \ref{fig:Qsi_Q0_regions}), these forcings do not exactly replicate reality. In ongoing work, we are extending this study by forcing the model with observed weather data from the Greenland Ice Sheet, and comparing to more specific observations of surface lowering, porosity and temperature. 

\section*{Acknowledgments}
Tilly Woods thanks Brian Wetton for valuable discussions about the enthalpy method.

\section*{Funding}
This work was supported by the UK Engineering and Physical Sciences Research Council [doctoral studentship to TW].

\appendix
\section{Details of numerical enthalpy method}
\label{sec:enthalpy_app}

We solve the enthalpy equation \eqref{eq:enthalpy_eqn} using a semi-implicit finite volume scheme in MATLAB. The flux $F$ is evaluated on cell edges, and the enthalpy $H$ is evaluated on cell centres. We include $\dot{h}$ and $F$ explicitly, $H$ in the advection term implicitly, and $T$ in the diffusion term semi-implictly, as follows
\begin{equation}
\label{eq:enthalpy_eqn_discrete}
\frac{H^{m+1}-H^{m}}{\Delta t} = - \dot{h}^{m} \frac{\partial H^{m+1}}{\partial Z} + k\frac{\partial^2 \big( T_{H}(H^{m+1}) \big)}{\partial Z^2} - \frac{\partial F^{m}}{\partial Z},
\end{equation}
where the superscripts $m$ and $m+1$ represent solutions at the current and next timestep. $T_{H}$ is a function that we use to express $T$ in terms of $H$, based on the transformation \eqref{eq:T_phi_from_H}.

We want to include $H$ in the diffusion term implicitly, but how $T$ is expressed in terms of $H$ depends on whether there is only ice, both ice and water, or only water (the three cases in \eqref{eq:T_phi_from_H}), which we do not know until we know the value of $H$. Similarly, the implementation of the surface boundary condition is different depending on whether or not there is surface melting, which, again, we do not know \textit{a priori}. We tackle these issues by using an `enthalpy flag' and a `melt flag', and iterating within each timestep to determine the appropriate values of these.

\subsection{Melt flag}
At each timestep, the surface could be in one of two `cases', which we label with a flag $f_\mathrm{melt}$: the melt case ($f_\mathrm{melt}=1$) or the no-melt case ($f_\mathrm{melt}=0$), with different surface boundary conditions being applied in each case. In the melt case, the surface is at the melting temperature, $T=T_\mathrm{m}$ at $Z=0$ is our boundary condition, and the surface energy balance \eqref{eq:SEB_simplified} tells us the surface melt rate $m_\mathrm{surf} > 0$. In the no-melt case, there is no surface melting ($m_\mathrm{surf}=0$), $T \leq T_\mathrm{m}$ at the surface, and the surface energy balance \eqref{eq:SEB_simplified} is the boundary condition. 

\subsection{Enthalpy flag}
Similarly, at each timestep, each point in the domain can be in one of the three states in \eqref{eq:T_phi_from_H}. We label these using a flag $\mathbf{f}_H$, referred to as the enthalpy flag:
\begin{equation}
\label{eq:enthalpy_flag}
\mathbf{f}_H = \begin{cases}
    1, & \mathrm{if \ there \ is \ only \ ice} \quad (\phi=0, \ T \leq T_\mathrm{m}, \ H \leq 0),\\
    2,  & \mathrm{if \ there \ is \ both \ ice \ and \ water} \quad (0<\phi<1, \ T = T_\mathrm{m}, \ 0<H< \rho \mathcal{L}), \\
    3,  & \mathrm{if \ there \ is \ only \ water} \quad (\phi=1, \ T \geq T_\mathrm{m}, \ H \geq \rho \mathcal{L}).
\end{cases}
\end{equation}
Note that $\mathbf{f}_H$ is a vector with its size equal to the number of grid cells.

\subsection{Iterations within timesteps}
To ensure we use the correct form of the surface boundary condition and the correct conversion from $T$ to $H$, we iterate the melt flag $f_\mathrm{melt}$ and enthalpy flag $\mathbf{f}_H$ as follows. At the start of each timestep, we assume the values of the flags to be the same as the previous timestep (e.g. if the surface was melting in the previous timestep then we assume it is in the new timestep too). Then within each timestep, we carry out 3 iterations. In each iteration, we
\begin{enumerate}
    \item Solve for $H^{m+1}$ using the values of $f_\mathrm{melt}$ and $\mathbf{f}_H$ from the previous iteration. If $f_\mathrm{melt}=1$, $T=T_\mathrm{m}$ is used as the surface boundary condition. If $f_\mathrm{melt}=0$, the surface energy balance \eqref{eq:SEB_simplified} is used with $m_\mathrm{surf}=0$. The value of $\mathbf{f}_H$ tells us how to express $T$ in the enthalpy equation \eqref{eq:enthalpy_eqn} in terms of $H$, following \eqref{eq:T_phi_from_H}.
    \item Update the melt flag if the values of $m_\mathrm{surf}$ and $T_\mathrm{surf}$ (surface temperature) in the solution contradict the assumed melt flag.
    \begin{itemize}
        \item Switch to $f_\mathrm{melt}=0$ if we assumed $f_\mathrm{melt}=1$ and calculated $m_\mathrm{surf} \leq 0$ (i.e. if we assumed there was surface melting but calculated none).
        \item Switch to $f_\mathrm{melt}=1$ if we assumed $f_\mathrm{melt}=0$ and calculated $T_\mathrm{surf} \ge T_\mathrm{m}$ (i.e. if we assumed there was no melting but calculated that the surface was above the melting temperature).
    \end{itemize}
    \item Update the enthalpy flag $\mathbf{f}_H$ if $T$ and $\phi$ in the solution contradict the assumed enthalpy flag.
    \begin{itemize}
        \item If we assumed only ice ($\mathbf{f}_H=1$) but found $T>T_\mathrm{m}$ ($H>0$), set $\mathbf{f}_H=2$ (try assuming there is both ice and water).
        \item If we assumed there is both ice and water ($\mathbf{f}_H=2$) but found $\phi < 0$ ($H<0$), set $\mathbf{f}_H=1$ (try assuming there is only ice).
        \item If we assumed there is both ice and water ($\mathbf{f}_H=2$) but found $\phi > 1$ ($H>\rho \mathcal{L}$), set $\mathbf{f}_H=3$ (try assuming there is only water).
        \item If we assumed only water ($\mathbf{f}_H=3$) but found $T<T_\mathrm{m}$ ($H<0$), set $\mathbf{f}_H=2$ (try assuming there is both ice and water).
    \end{itemize}
    \item If the assumed melt flag and the enthalpy flag at each point in the domain were `correct' (i.e. the values from the solution are equal to what we assumed), stop iterating.
\end{enumerate}

Note that the inequalities in step 2 are weak. This is to prevent the surface from getting `stuck' in a no-melt state or vice versa. Having a solution with $m_\mathrm{surf}=0$ and $T_\mathrm{surf}=T_\mathrm{m}$ is possible in both states, so having equality is not a contradiction. However, if $m_\mathrm{surf}=0$ is achieved having previously been in a state with $m_\mathrm{surf}>0$, it suggests that the surface will likely want to switch to being frozen in the next timestep, and not allowing this can cause issues.

\section{Parameter values}
\label{sec:parameters_app}
The parameter values used throughout this study are listed in Table \ref{tab:parameters_dimensional}.

\begin{table}[t]
\centering
\begin{tabular}{|l|c|c|c|}
\hline
Quantity & Symbol & Default value & Units\\
\hline
Density of ice & $\rho$ & 910 & kg m$^{-3}$ \\
Specific heat capacity of ice at 0$^\mathrm{o}$C &$c$ & 2097 & J kg$^{-1}$ K$^{-1}$ \\
Thermal conductivity& $k$ & 2.1 &  kg m$^{-3}$ K$^{-1}$\\
Melting temperature& $T_\mathrm{m}$ & 273.15 & K  \\
Latent heat of melting &$\mathcal{L}$ & 334000 & m$^2$ s$^{-2}$  \\
Albedo & $a$ & 0.6 & \\
Proportion of radiation absorbed at the surface$^{\mathrm{a}}$ & $\chi$  & 0.36&  \\
Extinction coefficient of internal radiation$^\mathrm{b}$ & $\kappa$ & 1.5  & m$^{-1}$ \\
Absorption coefficient$^\mathrm{c}$ &$\alpha$ &  0.2637  & m$^{-1}$ \\
Scattering coefficient$^\mathrm{c}$ & $r$ & 4.1338  & m$^{-1}$ \\
Incoming shortwave radiation flux & $Q_\mathrm{si}$ & various & W m$^{-2}$ \\
Turbulent heat and longwave radiation flux & $Q_0$ & various  & W m$^{-2}$ \\
Effective heat transfer coefficient$^\mathrm{d}$ & $\upsilon$ & 14.8 & W m$^{-2}$ K$^{-1}$ \\
Temperature at depth & $T_{\infty}$ & -10  & $^{\mathrm{o}}$C \\
\hline
\end{tabular}
\caption{\scriptsize Default values of the parameters in the model. Unless otherwise stated, these are the parameter values used. $^\mathrm{a}$Motivated by \citet{Hoffman2014}. $^\mathrm{b}$\citet{Taylor2005}. $^\mathrm{c}$Calculated from $a$, $\kappa$ and $\chi$ as in \citet{Woods2023}. $^\mathrm{d}$Calculated as $\nu = C + 4 \epsilon \sigma T_\mathrm{m}^3$ \citep{Woods2023}, where $C = 10.3$ W m$^{-2}$ K$^{-1}$ is a heat transfer coefficient \citep{Cuffey2010}, $\epsilon = 0.97$ is the surface emissivity \citep{Meyer2017}, $\sigma = 5.7 \times 10^{-8}$ W m$^{-2}$ K$^{-4}$ is the Stefan-Boltzmann constant, and $T_\mathrm{m} = 273.15$ K is the melting temperature of ice \citep{Cuffey2010}.
Many of the ice properties are taken from \citet{Cuffey2010}.}
\label{tab:parameters_dimensional}
\end{table}

\bibliography{references}

\end{document}